\documentclass{ws-procs975x65}

\begin{document}

\markboth{Vas\'uth, Mik\'oczi, Gergely}
{Orbital phase in inspiralling compact binaries}

\title{ORBITAL PHASE IN INSPIRALLING COMPACT BINARIES
\footnote{
Research supported by OTKA grants nos. T046939, TS044665, F049429 and the J\'{a}nos
Bolyai Fellowships of the Hungarian Academy of Sciences. M.V. and L.\'A.G. wish to
thank the organizers of the 11th Marcel Grossmann Meeting for support.} 
}

\author{M\'ATY\'AS VAS\'UTH$^\dagger$, BAL\'AZS MIK\'OCZI$^\ddagger$ and L\'ASZL\'O
\'A. GERGELY$^\dagger$}

\address{$^\dagger$KFKI Research Institute for Particle and Nuclear Physics \\ 
Budapest 114, P.O.Box 49, H-1525, Hungary \\
$^\ddagger$Departments of Theoretical and Experimental Physics, University of Szeged \\ 
D\'om t\'er 9, Szeged H-6720, Hungary \\
\email{vasuth@rmki.kfki.hu, mikoczi@titan.physx.u-szeged.hu, gergely@physx.u-szeged.hu}
}

\begin{abstract}
We derive the rate of change of the mean motion up to the second
post-Newtonian order for inspiralling compact binaries with spin, mass
quadrupole and magnetic dipole moments on eccentric orbits. We give this
result in terms of orbital elements. We also present the related orbital
phase for circular orbits.
\end{abstract}


\bodymatter

\section*{}
Observations by Earth-based gravitational wave observatories are under way
aiming to detect gravitational radiation. Upper limits from interferometer
data were already set on inspiral event rates for both binary neutron stars
\cite{LIGONS} and binaries of $3-20$ solar mass black holes.\cite{LIGOBH}
The parameters of spinning compact binaries can be estimated and alternative
theories of gravity can also be tested from these measurements.\cite{BBW}

An important characteristic of these binaries is the rate of decrease of the
orbital period $T$ due to the energy and angular momentum carried away by
gravitational waves. Here we give the radiative change of the mean motion 
$n=2\pi /T$ (for eccentric orbits). We also present the related change
occured in the orbital phase (for circular orbits). In both expressions we
include all known \textit{linear} perturbations for an isolated compact
binary. These are the post-Newtonian (PN), spin-orbit (SO), spin-spin (SS),
self spin (Self, quadratic in the single spins), quadrupole-monopole (QM)
and magnetic dipole-magnetic dipole (DD) contributions.

The expression of the radial period, defined as half of the time elapsed
between the turning points, emerges from generic considerations on the
perturbed Keplerian motion.\cite{Param,Param2} Collecting all linear
contributions the mean motion has the following~form 
\begin{equation}
n=\frac{\left( 2\mathcal{E}\right) ^{3/2}}{Gm}\left[ 1-(15-\eta )\frac{%
\mathcal{E}}{4c^{2}}\right] \ ,  \label{meanmot}
\end{equation}%
where $\eta =\mu /m$ is the ratio of the reduced mass $\mu $ to the total
mass $m$ of the binary system, and $\mathcal{E}=-E/\mu $ where $E$ is the
conserved energy. Remarkably there are no explicit spin, quadrupolar and
magnetic dipolar contributions in the functional form of the mean motion.
These however contribute implicitly to $n$ through $\mathcal{E}$. Since the
mean motion is a function of $E$ alone, its evolution can be computed as $%
\langle dn/dt\rangle =-1/\mu (dn/d\mathcal{E})\langle dE/dt\rangle $. All
linear contributions to the secular energy loss $\langle dE/dt\rangle $ due
to finite size effects are explicitly given in the literature\cite%
{SO,SS,QM,DD}, in terms of dynamical constants. The PN contribution is also
well-known.\cite{PN} Employing these we find the change of the mean motion: 
\begin{eqnarray}
\left\langle \frac{dn}{dt}\right\rangle _{N} &=&N_{0}\ , \\
\left\langle \frac{dn}{dt}\right\rangle _{PN}\!\! &=&-{\,}{\frac{{G\,m}}{56{{%
c^{2}a}}(1-e^{2})}}\left[ N_{1}+N_{2}\eta \right] \ , \\
\left\langle \frac{dn}{dt}\right\rangle _{SO} &=&-\frac{G^{1/2}}{%
2c^{2}a^{3/2}(1-e^{2})^{3/2}m^{1/2}}\left[ N_{3}S_{L}+N_{4}\Sigma _{L}\right]
\ , \\
\left\langle \frac{dn}{dt}\right\rangle _{Self}\!\! &=&\!\!\frac{1}{%
64c^{2}a^{2}(1-e^{2})^{2}}\!\!\sum_{i=1}^{2}\!\!\left(\!\frac{S_{i}}{m_{i}}\!\right)^{2}
\!\!\!\left[ N_{5}\sin ^{2}\!\kappa _{i}\cos 2(\psi
_{0}\!\!-\!\!\psi _{i})\!+\!N_{6}(6\!+\sin^{2}\!\kappa _{i}\!)\right] ,
\\
\left\langle \frac{dn}{dt}\right\rangle _{SS}{}\! &=&\!{}\!{}\frac{S_{1}S_{2}%
}{32c^{2}m\mu a^{2}(1-e^{2})^{2}}  \nonumber \\
&&\times {}\!\left[ N_{7}\sin \kappa _{1}\sin \kappa _{2}\cos 2(\psi
_{0}\!-\!\bar{\psi})+N_{8}\cos \kappa _{1}\cos \kappa _{2}+N_{9}\cos \gamma\right] \ , \\
\left\langle \frac{dn}{dt}\right\rangle _{QM} &=&\!\frac{m^{2}}{%
4a^{2}(1-e^{2})^{2}}\!\sum_{i=1}^{2}p_{i}\left[ N_{10}(2\!-\!3\sin
^{2}\!\kappa _{i})\!+\!N_{11}\sin^{2}\!\kappa _{i}\cos 2(\psi _{0}\!-\!\psi
_{i})\right] \ , \\
\left\langle \frac{dn}{dt}\right\rangle _{DD} &=&-\frac{d_{1}d_{2}}{2Gm\mu
a^{2}(1-e^{2})^{2}}(N_{10}\mathcal{A}_{0}+N_{11}\mathcal{B}_{0})\ ,
\label{dndt}
\end{eqnarray}%
where we have introduced the notations $S_{L}=(S_{1}\cos \kappa
_{1}+S_{2}\cos \kappa _{2})$, $\Sigma _{L}=\left[ \left( m_{2}/m_{1}\right)
S_{1}\cos \kappa _{1}+\left( m_{1}/m_{2}\right) S_{2}\cos \kappa _{2}\right] 
$ and 
\begin{equation}
N_{j}=\frac{G^{7/2}m^{5/2}\mu }{5c^{5}a^{11/2}(1-e^{2})^{7/2}}%
\sum_{i=0}^{3}n_{ij}e^{2i},\qquad j=0..11\ ,
\end{equation}%
Here $\kappa _{i}$ and $\psi _{i}$ are the polar and azimuthal angles of the 
$i^{\text{th}}$ spin vector and the numeric coefficients $n_{ij}$ are:
\[
\begin{tabular}{|l|l@{\hspace{8.5pt}}l@{\hspace{8.5pt}}l@{\hspace{8.5pt}}l@{\hspace{8.5pt}}l@{\hspace{8.5pt}}l
@{\hspace{8.5pt}}l@{\hspace{8.5pt}}l@{\hspace{8.5pt}}l@{\hspace{8.5pt}}l@{\hspace{8.5pt}}l@{\hspace{8.5pt}}l|}
\hline
{\footnotesize i} & {\footnotesize j=0} & {\footnotesize 1} & {\footnotesize 2} & {\footnotesize 3} & 
{\footnotesize 4} & {\footnotesize 5} & {\footnotesize 6} & {\footnotesize 7} & {\footnotesize 8} & 
{\footnotesize 9} & {\footnotesize 10} & {\footnotesize 11} \\ \hline\hline
{\footnotesize 0} & {\footnotesize 96} & {\footnotesize 28016} & {\footnotesize 9408} & {\footnotesize 2128} & 
{\footnotesize 1680} & {\footnotesize 0} & {\footnotesize 64} & {\footnotesize -3072} & {\footnotesize -194368} & 
{\footnotesize 65216} & {\footnotesize 2888} & {\footnotesize 288}\\ 
{\footnotesize 1} & {\footnotesize 292} & {\footnotesize 160248} & {\footnotesize 43120} & {\footnotesize 7936} & 
{\footnotesize 7924} & {\footnotesize 16} & {\footnotesize 608} & {\footnotesize 100112} & {\footnotesize -621536} & 
{\footnotesize 211232} & {\footnotesize 9660} & {\footnotesize -7924} \\ 
{\footnotesize 2} & {\footnotesize 37} & {\footnotesize 34650} & {\footnotesize 20916} & {\footnotesize 3510} & 
{\footnotesize 4224} & {\footnotesize 80} & {\footnotesize 552} & {\footnotesize 113248} & {\footnotesize -264792} & 
{\footnotesize 91944} & {\footnotesize 3897} & {\footnotesize -8570} \\ 
{\footnotesize 3} & {\footnotesize 0} & {\footnotesize -5501} & {\footnotesize -1036} & {\footnotesize 363} & 
{\footnotesize 291} & {\footnotesize 9} & {\footnotesize 36} & {\footnotesize 8937} & {\footnotesize -4500} & 
{\footnotesize 1740} & {\footnotesize 187} & {\footnotesize -464} \\ \hline
\end{tabular}
\]

The orbital elements $a,\ e$ were derived\cite{KMG} from the turning points
of the radial part of the perturbed motion cf. $r_{{}_{{}_{min}^{max}}}=a%
\left( 1\pm e\right) $. (In these variables $n=\left( Gm/a^{3}\right) ^{1/2}%
\left[ 1+\left( \eta -9\right) Gm/2ac^{2}\right] \!$ ).

For Keplerian motions the orbital frequency $\omega =n$. Due to the
perturbations, precessions occur in the plane of motion (like periastron
advance), and the plane of motion can also evolve. Therefore the relation of 
$\omega$ and $n$ becomes more complicated.\cite{WS} In the presence of the
PN, SO, SS, Self, QM and DD perturbations, for circular orbits ($e=0$) the
change in the orbital frequency due to gravitational radiation is:\cite{Kidder,MVG} 
\begin{eqnarray}
\left\langle \frac{d\omega }{dt}\right\rangle &=&\frac{96\eta m^{5/3}\omega
^{11/3}}{5}\Biggl[1-\left( \frac{743}{336}+\frac{11}{4}\eta \right) \left(
m\omega \right) ^{2/3}  \nonumber \\
&+&\left( 4\pi -\beta \right) m\omega +\Biggl(\frac{34103}{18144}+\frac{13661%
}{2016}\,\eta +\frac{59}{18}\,\eta ^{2}+\sigma \Biggr)\left( m\omega \right)
^{4/3}\Biggr]\ ,  \label{omegagadot}
\end{eqnarray}%
where $\sigma =\sigma _{SS}+\sigma _{Self}+\sigma _{QM}+\sigma _{DD}\ $and $%
\beta $, $\sigma _{S_{1}S_{2}}$, $\sigma _{Self}$, $\sigma _{QM}$, $\sigma
_{DD}$ are the spin-orbit, spin-spin, self-interaction spin,
quadrupole-monopole and magnetic dipole-dipole parameters.\cite{MVG} For
completeness we have added the 2PN and tail contributions\cite{Blanchet95}
and we note that higher order contributions are also known.\cite{3.5PN,PNSO}

In terms of the dimensionless time variable $\tau =\eta (t_{c}-t)/5m$,
defined in terms of the time $(t_{c}-t)$ left until the final coalescence,
the accumulated orbital phase is $\phi =\phi _{c}-\left( 5m/\eta \right)
\int \omega (\tau )d\tau $, where $\phi _{c}$ is an integration constant. To
second post-Newtonian order: 
\begin{eqnarray}
\phi &=&\phi _{c}-\frac{1}{\eta }\Biggl\{\tau ^{5/8}+\left( \frac{3715}{8064}%
+\frac{55}{96}\eta \right) \tau ^{3/8}+\frac{3}{4}\left( \frac{\beta }{4}%
-\pi \right) \tau ^{1/4}  \nonumber \\
&&+\Biggl(\frac{9275495}{14450688}+\frac{284875}{258048}\eta +\frac{1855}{%
2048}\eta ^{2}-\frac{15\sigma }{64}\Biggr)\ \tau ^{1/8}\Biggr\}\ .
\end{eqnarray}
The modification induced by the finite size effects SS, Self, QM and DD are
all encoded in $\sigma $, while the SO contribution is in $\beta$.


\begin{thebibliography}{99}
\bibitem{LIGONS} B. Abbott et al, Phys. Rev. D \textbf{69}, 122001 (2004).

\bibitem{LIGOBH} E. Messaritaki, Class. Quantum Grav. \textbf{22}, S1119
(2005).

\bibitem{BBW} E. Berti, A. Buonanno, and C.~M. Will, Phys. Rev. D \textbf{71}%
, 084025 (2005).

\bibitem{Param} L.~\'{A}. Gergely, Z. Perj\'{e}s, and M. Vas\'{u}th,
Astrophys. J. Suppl. \textbf{126}, 79 (2000).

\bibitem{Param2} L.~\'{A}. Gergely, Z. Keresztes, and B. Mik\'{o}czi,
Astrophys. J. Suppl. \textbf{167}, 286 (2006).

\bibitem{SO} L. \'{A}. Gergely, Z. I. Perj\'{e}s, and M. Vas\'{u}th, Phys.
Rev. D \textbf{58}, 124001 (1998).

\bibitem{SS} L. \'{A}. Gergely, Phys. Rev. D \textbf{61}, 024035 (2000).

\bibitem{QM} L. \'{A}. Gergely, Z. Keresztes, Phys. Rev. D \textbf{67},
024020 (2003).

\bibitem{DD} M. Vas\'{u}th,\! Z. Keresztes,\! A. Mih\'{a}ly, and L. \'{A}.
Gergely, Phys. Rev. D \textbf{68}, 124006 (2003).

\bibitem{PN} A. Gopakumar, Bala R. Iyer Phys. Rev. D \textbf{56}, 7708
(1997).

\bibitem{KMG} Z. Keresztes, B. Mik\'{o}czi and L.~\'{A}. Gergely Phys.Rev. D 
\textbf{72}, 104022 (2005).

\bibitem{WS} G. Sch\"{a}fer, N. Wex, Phys. Lett A \textbf{174}, 196 (1993),
erratum: \textbf{177}, 461 (1993).

\bibitem{Kidder} L. Kidder, Phys. Rev. D \textbf{52}, 821 (1995).

\bibitem{MVG} B. Mik\'{o}czi, M. Vas\'{u}th, and L.~\'{A}. Gergely Phys.
Rev. D \textbf{71}, 124043 (2005).

\bibitem{Blanchet95} L. Blanchet, Phys. Rev. D \textbf{54}, 1417 (1996).

\bibitem{3.5PN} L. Blanchet, G. Faye, B. R. Iyer, and B. Joguet Phys. Rev. D 
\textbf{65}, 061501 (2002); Erratum ibid. D \textbf{71}, 129902 (2005).

\bibitem{PNSO} L. Blanchet, A. Buonanno, and G. Faye Phys. Rev. D \textbf{74}%
, 104034 (2006); Erratum ibid. D \textbf{75}, 049903 (2007).
\end{thebibliography}
\end{document}